\documentclass[11pt]{article}

\date{}

\usepackage[utf8]{inputenc} 
\usepackage{hyperref}       
\usepackage{url}            
\usepackage{booktabs}       
\usepackage{amsfonts}       
\usepackage{nicefrac}       
\usepackage{microtype}      
\usepackage{xspace}

\usepackage{comment}
\usepackage{amssymb}
\usepackage{amsmath}
\usepackage{amsthm}
\usepackage{graphicx}
\usepackage{rotating}

\usepackage{amssymb}
\usepackage{autobreak}
\usepackage{mathtools}
\usepackage{xcolor}
\usepackage{bbm}
\usepackage{algorithm}
\usepackage{algorithmic}
\usepackage{authblk}

\graphicspath{ {figs/} }

\usepackage[ 
backend=biber, 
natbib=true,
style=apa,
sorting=ynt]{biblatex}
\makeatletter

\title{Data-driven Accelerogram Synthesis using Deep Generative Models}

\author[1]{Manuel A. Florez}
\author[2]{Michaelangelo Caporale}
\author[3]{Pakpoom Buabthong}
\author[1]{Zachary E. Ross}
\author[2]{Domniki Asimaki}
\author[1]{Men-Andrin Meier}

\affil[1]{Seismological Laboratory, California Institute of Technology, Pasadena, CA}
\affil[2]{Division of Engineering and Applied Sciences, California Institute of Technology, Pasadena, CA}
\affil[3]{Division of Chemistry and Chemical Engineering, California Institute of Technology, Pasadena, CA}

\DeclareMathOperator{\Pg}{\mathbb{P}_g}
\DeclareMathOperator{\Preal}{\mathbb{P}_r}
\DeclareMathOperator{\E}{\mathbb{E}}
\DeclareMathOperator{\R}{\mathbb{R}}
\DeclareMathOperator{\Px}{\mathbb{P}_{x'}}
\DeclareMathOperator{\vc}{\mathbf{v}}
\DeclareMathOperator{\wa}{\textbf{w}}

\DeclareUnicodeCharacter{2217}{*}
\begin{document}

\maketitle

\begin{abstract}

Robust estimation of ground motions generated by scenario earthquakes is critical for many engineering applications. We leverage recent advances in Generative Adversarial Networks (GANs) to develop a new framework for synthesizing earthquake acceleration time histories. Our approach extends the Wasserstein GAN formulation to allow for the generation of ground-motions conditioned on a set of continuous physical variables. Our model is trained to approximate the intrinsic probability distribution of a massive set of strong-motion recordings from Japan. We show that the trained generator model can synthesize realistic 3-Component accelerograms conditioned on magnitude, distance, and $V_{s30}$. Our model captures the expected statistical features of the acceleration spectra and waveform envelopes. The output seismograms display clear P and S-wave arrivals with the appropriate energy content and relative onset timing. The synthesized Peak Ground Acceleration (PGA) estimates are also consistent with observations. We develop a set of metrics that allow us to assess the training process's stability and tune model hyperparameters. We further show that the trained generator network can interpolate to conditions where no earthquake ground motion recordings exist. Our approach allows the on-demand synthesis of accelerograms for engineering purposes.

\end{abstract}

\section{Introduction}

Ground motion time histories are a critical input for many engineering design tasks \citep{heaton_estimating_1986}. Under specific circumstances, modern construction codes make it compulsory to perform time-history analysis of a building's structural response \citep{bommer_use_2004}. Synthesizing realistic acceleration time-series remains a formidable challenge \citep{graves_broadband_2010, douglas_survey_2008}; practicing engineers still use past earthquakes recordings to represent the expected ground motions that would be generated by scenario events \citep{hancock_numbers_2008,hancock_improved_2006}. The assumption that future shaking will be similar to that observed in the past is well-grounded and typically valid.  Nevertheless, careful scaling and selection of records, as well as ease of access to databanks, are required to apply this technique successfully. Given the recent explosion in seismological data collection and the many exciting developments in modern machine learning, the idea that an artificial intelligence system could provide on-demand, accurate and realistic ground motion time histories for engineering purposes is not farfetched. 

While advances in our understanding of the complex rupture and wave propagation processes that cause the ground motions observed at the surface make the prospect of numerically simulating realistic strong motions increasingly likely, fundamental limitations remain. The exponential growth in computing power and the development of efficient and accurate numerical methods \citep{graves_simulating_1996,komatitsch_spectral_1998}, make it possible to model wave propagation through 3-D heterogeneous media. If an accurate velocity model is available, elastodynamic Green's functions for frequencies as high as 1 Hz can be calculated with relative ease. Once a reasonable approximation of the Green's function is obtained, the rupture process has to be modeled. A kinematic source description is often used \citep{herrero_kinematic_1994}; the hypothetical fault is divided into patches or sub-faults, and the relevant physical parameters (e.g., slip, rupture velocity, rise time) must be either known or assumed for each sub-fault \citep{graves_broadband_2010}.  A fully dynamic rupture simulation could capture the fundamental physical processes that drive faulting, and thus it would require a smaller number of well-constrained parameters; it is a promising alternative, actively researched, but still unavailable for any practical purposes \citep{mena_pseudodynamic_2012}. Current simulation methods account for large structures  \citep{ma_effects_2007},  such as sedimentary basins, but do not capture small scale heterogeneities responsible for local site effects \citep{graves_broadband_2010} and seismogram coda, which control the shape of the Fourier Amplitude Spectrum of accelerograms. Our lack of detailed knowledge of the earth's structure also poses a significant challenge; 3-D velocity models are notoriously difficult to constrain and only available for a handful of regions.

When the computational cost of deterministic simulations was prohibitive and observational datasets were small, sparse, and difficult to access \citep{douglas_survey_2008}, stochastic modeling methods were the only viable alternative. Initially developed for engineering purposes \citep{kaul_spectrum-consistent_1978, gasparini_simulated_1976, naeim_use_1995}. In their most basic form, white Gaussian noise is windowed and filtered, modified in the frequency domain, and then transformed back to the time domain, where it is multiplied by an envelope function \citep{naeim_use_1995}. The process is iterative, and the goal used to be that of matching a design response spectrum \citep{naeim_use_1995}. Pioneering work by \cite{boore_stochastic_1983,boore_simulation_2003-1} extended and formalized this approach for point sources by considering simple physical models \citep{boore_stochastic_1983}. The source spectrum is assumed to have a random phase \citep{hanks_character_1981}, but instead of using ad hoc manipulations, spectral amplitudes are modified to approximate the acceleration spectrum proposed by \citet{brune_tectonic_1970}, simplified theoretical representations of Path and Site effects are also needed. When transforming back to the time domain, an appropriate envelope function, reflecting source duration, must be carefully chosen \citep{boore_simulation_2003-1}. This approach can readily be extended to finite faults and does provide reasonable high-frequency approximations for single-phase arrivals \citep{herrero_kinematic_1994}. Unfortunately, it can not generate coherent 3-Component  waveforms with multiple arrivals and coda typical of real earthquakes \citep{douglas_survey_2008}; it completely ignores phase effects, and it does not accurately represent long period motions \citep{graves_broadband_2010}. Given these limitations, substantially more involved hybrid methods exist, in which low-frequency waveforms are simulated deterministically, while high-frequency effects are modeled stochastically \citep{graves_broadband_2010}.

We present a novel data-driven framework for synthesizing ground motion time histories. We train a generator model to learn an optimal probabilistic representation of observed acceleration time-series directly from a strong-motion dataset. In section 2, we review some of the concepts necessary to understand or approach.  Section 3 presents the details of the technique and model architecture. Section 4 describes the training dataset and our choice of input variables.  The result's section shows that our approach generalizes well, even when no data is available for specific event-station distance and earthquake magnitude ranges, and it also introduces a metric for assessing the training process's convergence; we focus on displaying examples and assessing the quality of our model for distances and magnitudes relevant to engineering applications. Finally, we discuss the advantages and potential limitations of our framework.

\section{Background on GANs}

Generative models are a class of statistical models that attempt to capture the underlying probability distribution of a dataset. In particular, they are trained to produce data that looks as if it was sampled from the original training set \citep{bengio_better_2013,alain_what_2014}. 
Data generation is harder than any classification or prediction task \citep{bengio_representation_2013}. A useful generative model has to adequately capture most of the correlations in data space. When learning to produce images of human faces, for example, a model must learn a data boundary: a human face only has two eyes, but it also has to position them below the forehead and place a nose in between. If the task is to synthesize seismograms, the S wave must come after the P wave, but more importantly, the statistical distribution of the data in frequency space should also be captured.

Generative Adversarial Networks (GANs) are state-of-the-art generative models \citep{goodfellow_generative_2014}. Advances in architecture and training techniques have enabled GANs to synthesize high-resolution realistic-looking images of human faces \citep{karras_progressive_2018}, audio \citep{donahue_adversarial_2018} and even video sequences \citep{saito_tganv2_2018}. GANs are built using two networks that are trained simultaneously \citep{goodfellow_generative_2014}: a discriminator network $D$ and a generator network $G$. The discriminator is a binary classification model trained to determine whether its input came from the real data distribution $\Preal$ or it was produced by $G$. The generator maps samples from a Gaussian distribution into samples coming from a new implicitly defined distribution $\Pg$. The goal of the process is to make $\Pg$ as close to $\Preal$ as possible.  

\begin{figure}[htp]
    \centering
    \includegraphics[width=\textwidth]{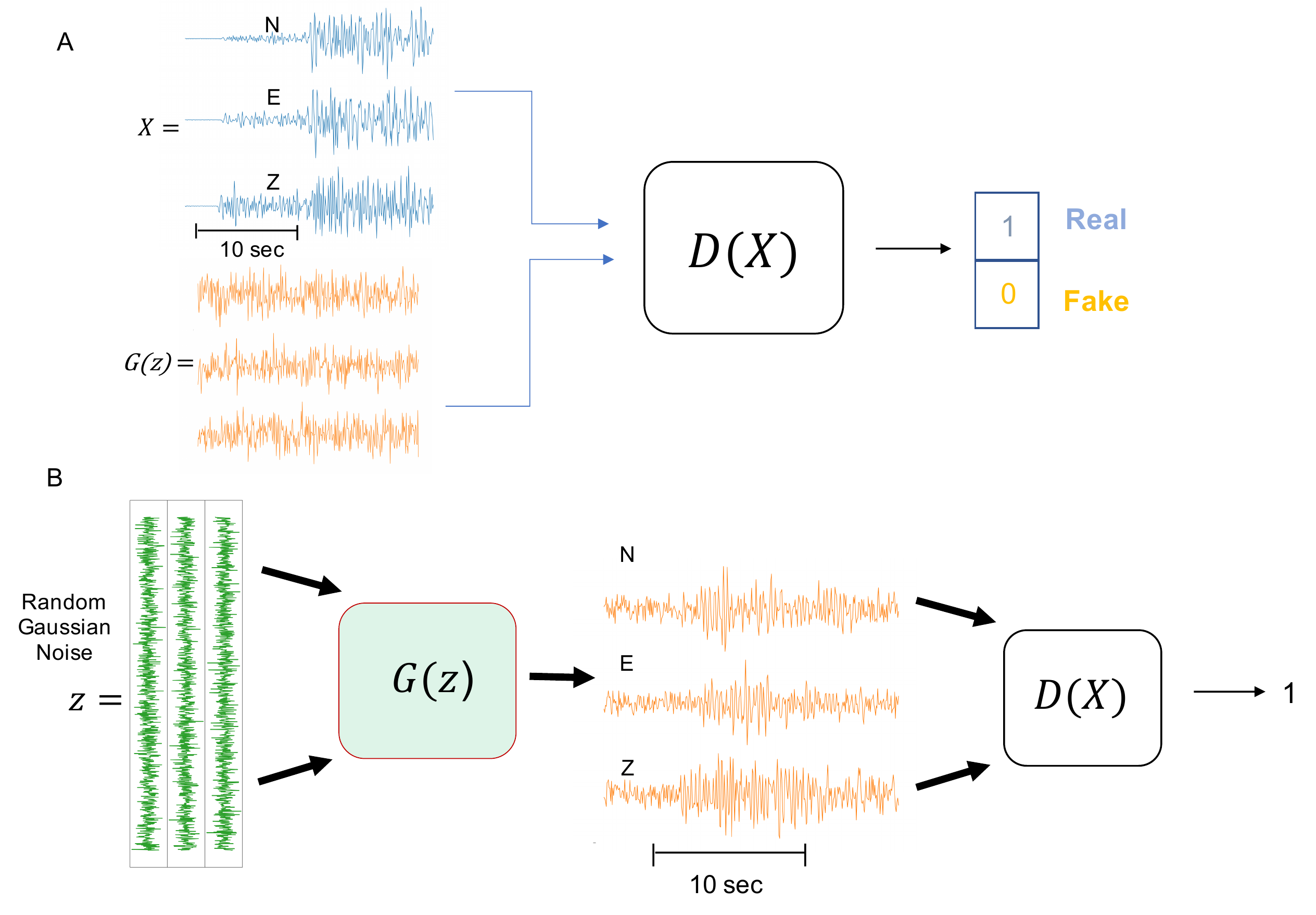}
    \caption{Schematic of a basic GAN architecture. $D$ represents the discriminator model; $G$ represents the generator model.}
    \label{fig:gan_basic}
\end{figure}

During training, the discriminator model is continuously refined; at each iteration, $D$ is shown samples generated by $G$, which are labeled as $0$, or "fake" and data sampled from $\Preal$, labeled with a $1$, or coming from the "real" distribution. The generator, on the other hand, is constantly optimized with respect to $D$, such that the samples it produces are classified as "real" by the discriminator \ref{fig:gan_basic}. Note that when the generator optimization step takes place, the weights of $D$ are kept fixed.

Despite the many successful implementations, issues such as instability and mode collapse make conventional GANs notoriously difficult to train \citep{salimans_improved_2016,arjovsky_towards_2017}. A given application might require substantial architectural refinement and hyperparameter tuning  \citep{radford_unsupervised_2015}. \citet{li_machine_2018} showed that, in principle, a GAN could be trained to generate synthetic seismograms. Unfortunately, their model was only able to capture the first 4 seconds of the P wave arrival. \citet{wang_seismogen_2019} also showed that waveforms synthesized using a GAN are useful for data augmentation and that an enhanced training set can improve earthquake detectability. However, finding a generative model able to synthesize realistic earthquake ground motion time-series remains an open problem.

\section{Methods}

\subsection{Wasserstein GANs}

Most of the work on GANs was focused on finding architectures to bring stability into the training process until the work of \citet{arjovsky_wasserstein_2017} provided the necessary theoretical tools to understand adversarial training. Conventional GANs, as introduced by \citet{goodfellow_generative_2014}, attempt to minimize the Jensen-Shannon (JS) divergence between the data distribution $\Preal$ and the distribution $\Pg$, implicitely defined by the generator model. Informally, the training instability arises from the fact that when $\Preal$ and $\Pg$ do not overlap, the JS divergence between them is not differentiable everywhere \citep{arjovsky_towards_2017}. Furthermore,  the discriminator can not be trained to optimality, as this would lead to increasingly small gradients \citet{arjovsky_wasserstein_2017}.

Wasserstein Generative Adversarial Networks (WGANs) overcome many of the limitations mentioned earlier by using the Earth Mover's (EM) distance as a measure of similarity between $\Pg$ and $\Preal$ \citep{arjovsky_wasserstein_2017}. Intuitively, the EM or Wasserstein-1 distance $W(\Pg,\Preal)$ is the minimum cost of transporting the probability mass necessary to transform $\Pg$ into the target distribution $\Preal$, where the cost is defined as mass times distance \citep{villani_optimal_2009}. In the Wasserstein formulation the discriminator is trained to solve the fallowing optimization problem:

\begin{equation} \label{eq:wdisc_optim}
\max_{D} \  \E_{x \sim \Preal}[D(x)]-\E_{z \sim p}[D(G(z))]
\end{equation}

Where $D(x) \in \R$ and $z \sim p$ implies that $z$ is sampled from a Gaussian distribution $p$. This is equivalent to finding $W(\Pg,\Preal)$ using the Kantorovich-Rubinstein duality \citep{villani_optimal_2009}. The discriminator $D(x)$ is no longer a binary classifier, it now acts as a critic, it outputs a real score that approximates the EM distance between $\Pg$ and $\Preal$. The optimization problem in (\ref{eq:wdisc_optim}) has an important constraint, $D$ must be a 1-Lipschitz function, that is \citep{arjovsky_wasserstein_2017}:

\begin{equation} \label{eq:Lip1}
\frac{\| D(x)-D(y) \|_2}{\| x-y \|_2} \leq 1
\end{equation}

For any input pair $x,y$, which means that the gradient of the discriminator with respect to its inputs is bounded, its $l_2$-norm can never exceed $1$. Weight clipping is the most straight forward way to impose (\ref{eq:Lip1}); unfortunately, it often leads to a coarse approximation of $\Preal$ and to vanishing or exploding gradients when the clipping threshold is not carefully chosen.

We follow \citet{gulrajani_improved_2017} and enforce the 1-Lipschitz constraint by adding a regularization term to the discriminator objective function:

\begin{equation} \label{eq:loss_disc}
L_{D} =  \E_{z \sim p}[ D(G(z)) ] - \E_{x \sim \Preal}[D(x)]+
\lambda \E_{x' \sim \Px}[(\| \nabla_{x'}D(x') \|_2-1)^2]
\end{equation}

Where $\lambda$ is a constant, and $x'$ is uniformly sampled along straight lines connecting points in $\Preal$ and $\Pg$. Imposing the 1-Lipschitz constraint via sampling does not guarantee that it will be satisfied everywhere; however, it gives good results \citep{gulrajani_improved_2017}, as it encourages the discriminator gradient norm to move towards $1$ in regions relevant to the problem, where gradients provide critical information for subsequent model updates. The generator is adversarially trained by minimizing the objective function:

\begin{equation} \label{eq:loss_gen}
L_{G} =  -\E_{z \sim p}[ D(G(z)) ]
\end{equation}

As in the original GAN formulation, the generator weights are optimized with respect to the discriminator, and both models are trained simultaneously. The 1-Lipschitz constraint on $D$ means that it can and should be trained to optimality, so it is useful to perform a few optimization steps, $N_c$, on the discriminator before the generator model is updated.

\subsection{Conditional WGAN}
We seek a model that can synthesize 3-Component accelerograms conditioned on a set of continuous input variables $\vc=(v_1,v_2, ..., v_k) $. Inspired by the success of Conditional Generative Adversarial Networks (cGAN) in many image generation tasks \citep{mirza_conditional_2014}, such as image to image translation \citep{isola_image--image_2017}, image super-resolution \citep{ledig_photo-realistic_2017} and text to image synthesis \citep{reed_generative_2016}; we present an extension of the Wasserstein GAN formulation for the task of data generation using continuous conditional variables.

\begin{figure}[htp]
    \centering
    \includegraphics[width=\textwidth]{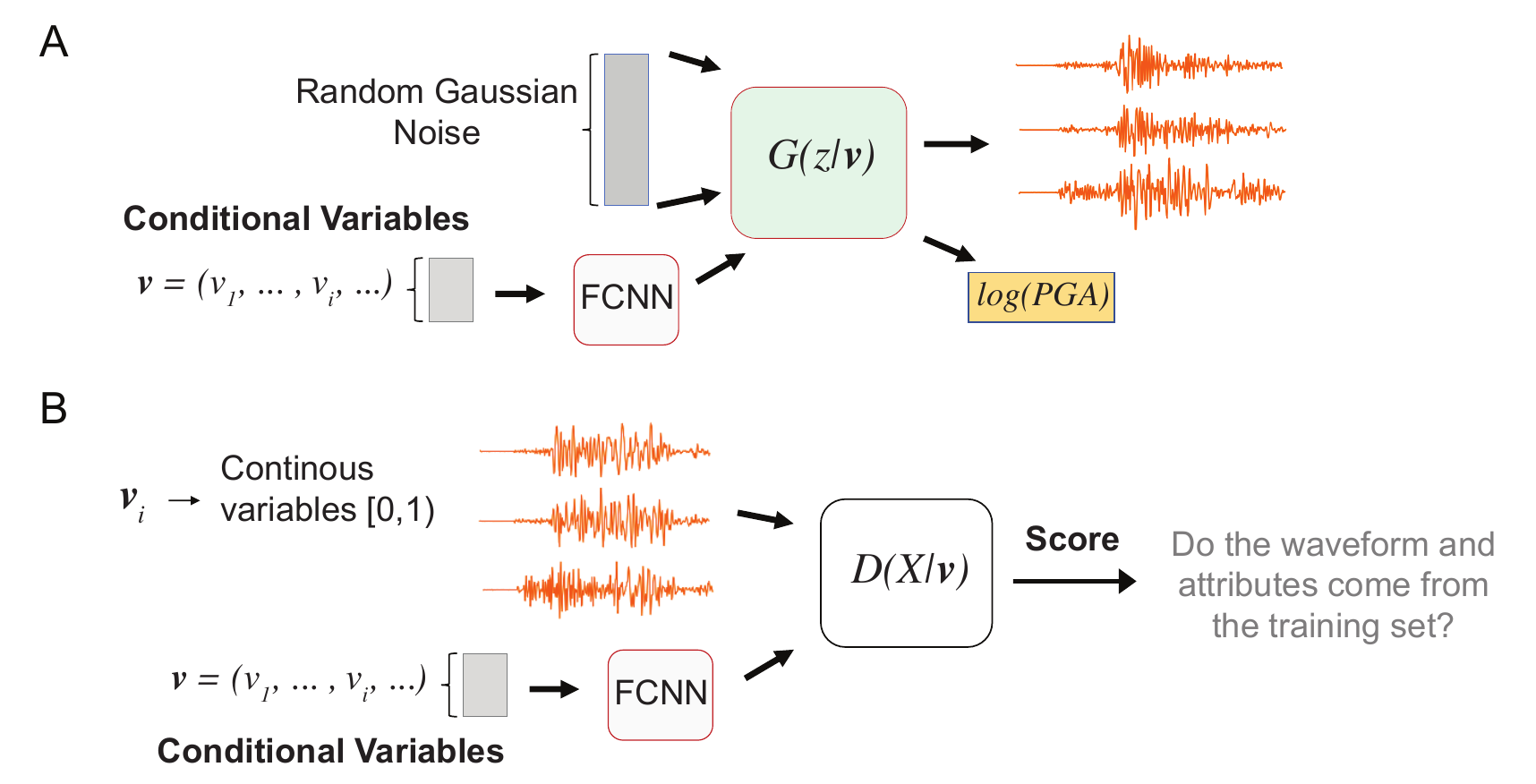}
    \caption{(A) Conditional Generator Model (B) Conditional Discriminator Model }
    \label{fig:cwgan}
\end{figure}

We construct a generator model to map normally distributed random noise $z$ and a set of continuous conditional variables $\vc$ into an accelerogram $\mathbf{w}$, $G: \{\vc, z \} \rightarrow w $ (figure \ref{fig:cwgan}). The mapping implicitly defines a conditional probability distribution $\Pg(\mathbf{w}|\vc)$. The discriminator model also learns a mapping, $G: \{\vc, w \} \rightarrow \R $, that approximates how close the distribution defined by $G$ is to the real conditional data distribution. Once more, the discriminator can be understood as a critic, trained to assess if the pair $\{\vc, \mathbf{w} \}$ was sampled from the real distribution. 

We empirically found that it was essential to use a separate embedding network for each of the $k$ conditional variables (figure \ref{fig:cwgan}). We designed each embedding as a four-layer fully connected neural network (FCNN) to map a continuous variable $v_i$ input to a higher dimensional representation. Both $D$ and $G$ had different embedding networks, whose weights were optimized during training. It is not clear why this is so critical, but probably, it is because learning is more effective when sparse high dimension data representations are used \citep{bengio_better_2013}.

It was also important to train the discriminator to optimality, which was achieved by using a relatively large number of critic iterations, $N_c$. We found that $N_c \leq 10$ was necessary to bring stability to the training process. A value of $5$ is typically used for unconditional image generation \citep{arjovsky_wasserstein_2017,gulrajani_improved_2017}, approximating the EM distance between conditional distributions might be a harder problem.

\begin{figure}[htp]
    \centering
    \includegraphics[width=\textwidth]{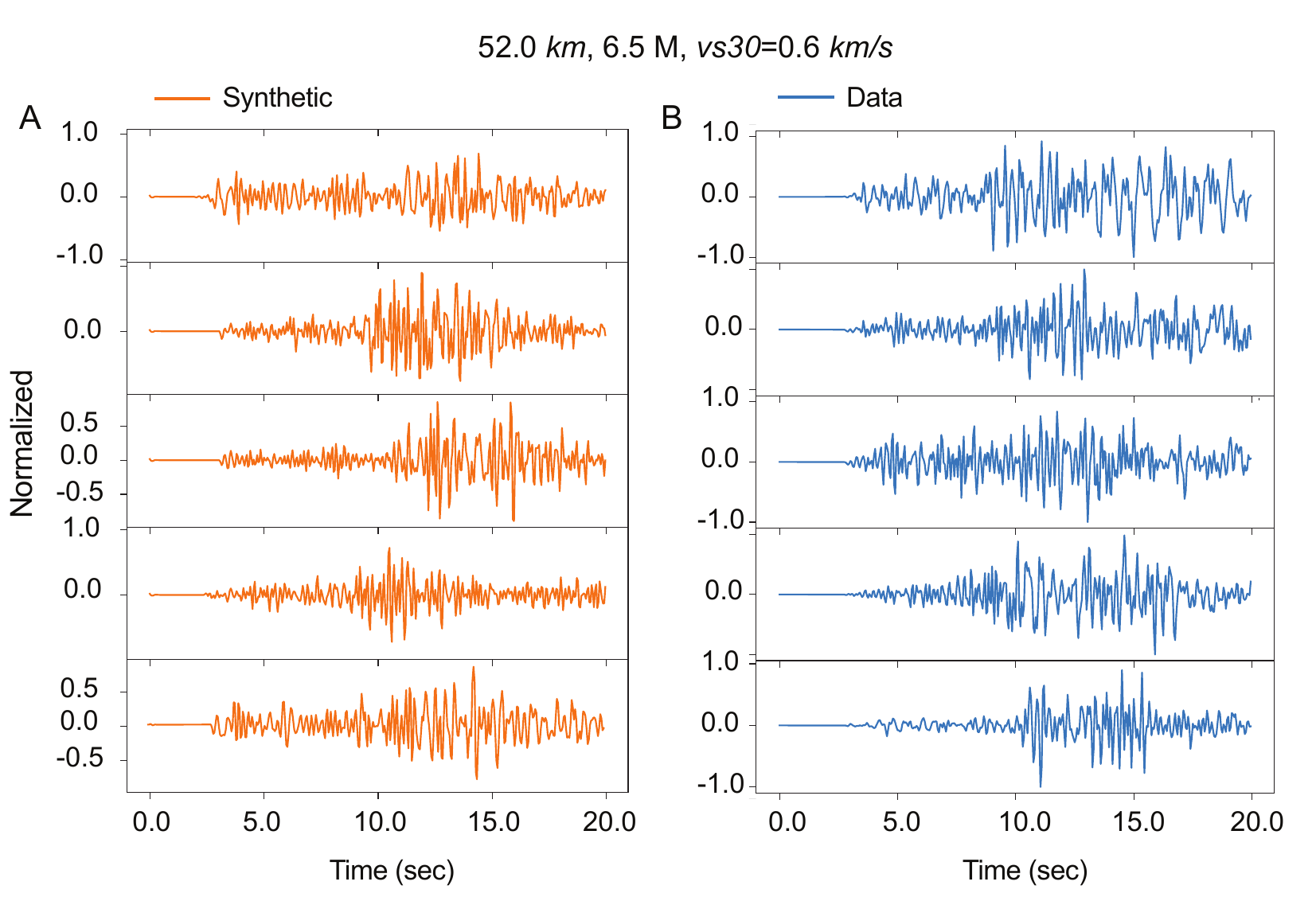}
    \caption{(A) Randomly sampled synthetic z-component accelerograms (orange), generated using $distance=52.0 km$, $M=6.5$ and $Vs30=0.6 km/s$ as conditional variables. (B) For comparison five real z-component accelerograms, randomly sampled from the bin:  $45.0-59.0 km$, $6.4-6.6 M$ and $0.4-0.8 km/s$.}
    \label{fig:cwaves}
\end{figure}

\subsection{Model Architecture}
The overall architecture follows the guidelines presented in \citep{radford_unsupervised_2015}, which have become standard practice when assembling deep convolutional GANs for image generation. The discriminator is implemented as a series of strided convolutions that progressively reduce the dimensionality of the input data, while increasing its depth, as the features relevant for classification are extracted.  The generator starts from a gaussian noise vector that is progressively upsampled until an output with the same shape as the discriminator's input is produced. In the formulation of \citet{radford_unsupervised_2015} upsampling is performed by a series of transpose convolutional layers, but we found this approach to be ineffective for synthesizing seismograms. Instead, we used a nearest-neighbor resize (NNR), followed by three successive convolutions \citep{odena_deconvolution_2016}. The NNR layer doubles the size of its input along the time dimension; the convolutional layers apply their filters while preserving their input's dimension along the time axis.

In our architecture, the generator model is given a three-channel random normal vector of length 100 as input, the higher dimensional representation of each conditional variable is concatenated and added as an additional channel. We feed the result into two fully connected layers with 128 units each and then into two upsampling layers implemented as described earlier, where each convolution has a filter of size 5 and a stride of 1. All layers are followed by batch normalization \citep{ioffe_batch_2015} and a ReLU activation function, except for the last one, where no normalization is applied, and $\tanh$ is used as activation. $G$ outputs 20 second long, 3-component waveforms, sampled at $20 Hz$.

The discriminator model takes a 3-component waveform, either real or synthesized by $G$, and the set of continuous conditional variables as inputs. After each variable is passed through its corresponding embedding network, its high dimensional representation is added as an extra channel to the waveform.  Four convolutional layers are followed by four fully connected layers. We apply Leaky ReLU $(\alpha=0.2)$ activation functions and batch normalization, except for the input layer, which omits the normalization, and the output layer, which omits both. With no activation, the output yields raw non-normalized values. In this manner, $D$ acts as a critic that scores its input, as opposed to a typical discriminator which assigns probabilistic binary labels.

\section{Data and Training}
\label{section:data}
We assemble a set of $260,764$ strong-motion recordings from Japanese seismograph networks K-NET and KiK-Net, corresponding to $6125$ earthquakes. We only use ground surface stations. We focus on signals useful for ground motion prediction and engineering applications \citep{douglas_survey_2008,power_overview_2008} by selecting events with magnitudes between 4.0 and 7.5 at event-station distances between $0$ and $280 km$. 

We also collect available values of $vs30$ at each station. $vs30$ is defined as the shear wave speed averaged to a depth of 30m  directly below the ground.  It is the best-known proxy for the response of a site to an earthquake \citep{borcherdt_vs30_2012}, and it is used in building codes worldwide to separate sites into different categories for design purposes.

\begin{figure}[htp]
    \centering
    \includegraphics[width=\textwidth]{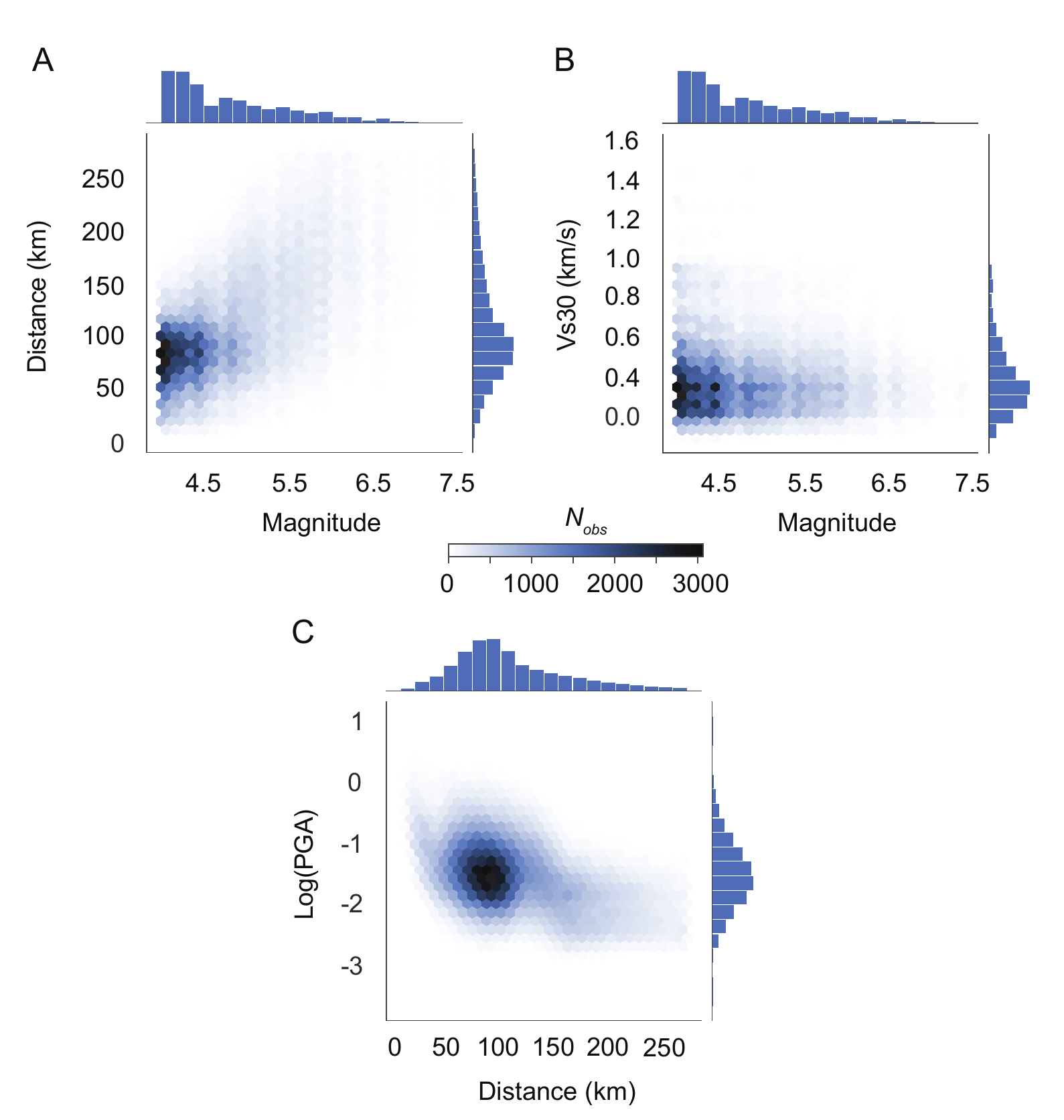}
    \caption{ Cross-plots with conditional variable distributions. Hexagonal bins are used to visualize  correlations between variables. Each bin is color-coded according to the number of observations that fall within the hexagon. (A) Event-Station distance and Magnitude. (B) Magnitude and $Vs30$. (C) $\log_{10} (PGA)$ and Event-Station Distance. }
    \label{fig:data}
\end{figure}

Because of their relevance to ground motion prediction, and their high correlation  with essential features of the ground motions generated at the surface, we select the following physical parameters as the conditional variables for our model (figure (\ref{fig:data})):
\begin{itemize}
    \item Event-Station Distance
    \item Earthquake Magnitude
    \item Vs30 at the recording station
\end{itemize}

Here we have focused on a minimal set of physical parameters whose impact on the observed seismograms can easily be quantified. These three variables are often the dominant terms in ground motion prediction equations \citep{douglas_survey_2008}, which makes our approach a useful supplement, or alternative, to current prediction methodologies.

Before training, we further curate our dataset. The original ground motion recordings are sampled at $100 Hz$; we downsample them to $20 Hz$ and select 20-second windows that start 2.5 seconds before the P wave onset. The maximum Peak Ground Acceleration (PGA) in our dataset is $1.2 g$. We also remove stations with less than 20 recorded accelerograms, as this could be indicative of potential instrumental issues.

We normalize each input channel to take values between $[-1,1]$ \citep{radford_unsupervised_2015}. The log of the scaling factors is concatenated to the input accelerogram. Thus, we train our model to synthesize 3-component, 400 sample seismograms, along with their associated normalization factors. We use the Adam stochastic optimization algorithm and train the discriminator $N_c=10$ iterations for each generator optimization step. We let the process run for at least $160$ epochs and select the best model by using the criteria discussed in section \ref{section:model_sel} as well as visual inspection of averaged statistical aspects of the generated waveforms.

\section{Results}

\subsection{Model Validation in Time and Frequency Domains}

Our generator model is stochastic; our goal is not to provide a one-to-one correspondence between a set of variables and seismograms, or ground motion intensity measures, as in a typical regression problem. Instead, by approximating the intrinsic probability distribution of a regional ground motion dataset, we have built a model that allows us to sample 3-Component acceleration time-series vectors, $ \wa(t)$, conditioned on a set of physically meaningful variables. As such, we can only assess the quality of our model in a statistical sense.  Because we seek to synthesize realistically looking seismograms adequate for engineering applications, our generator model must perform well in both frequency and time domains. We evaluate our model in the frequency domain (FD) by comparing the average Fourier amplitudes of real and synthetic seismograms. The average spectra of a set of $N$ accelerograms in log-space is given by:

\begin{equation} \label{eq:spec_ave}
\overline{A}(f) = \frac{1}{N} \sum_{k=1}^{N}\log(A_{k}(f))
\end{equation}

Where $A_{k}(f)$ is the norm of the 3-Component vector of Fourier Amplitudes for accelerogram $k$ at frequency $f$. We use a multitaper estimation technique to compute the spectrum of each accelerogram component \citep{prieto_fortran_2009}.

In the time domain (TD), we use average acceleration envelopes as our evaluation metric. The envelope of a seismogram $k$ is defined as $ \| \wa_{k}(t_i) \| $. As before, we compute averages in log-space:

\begin{equation} \label{eq:wa_ave}
\overline{W}(t_i) = \frac{1}{N} \sum_{k=1}^{N}\log( \| \wa_{k}(t_i) \|)
\end{equation}

\begin{figure}[htp]
    \centering
    \includegraphics[width=\textwidth]{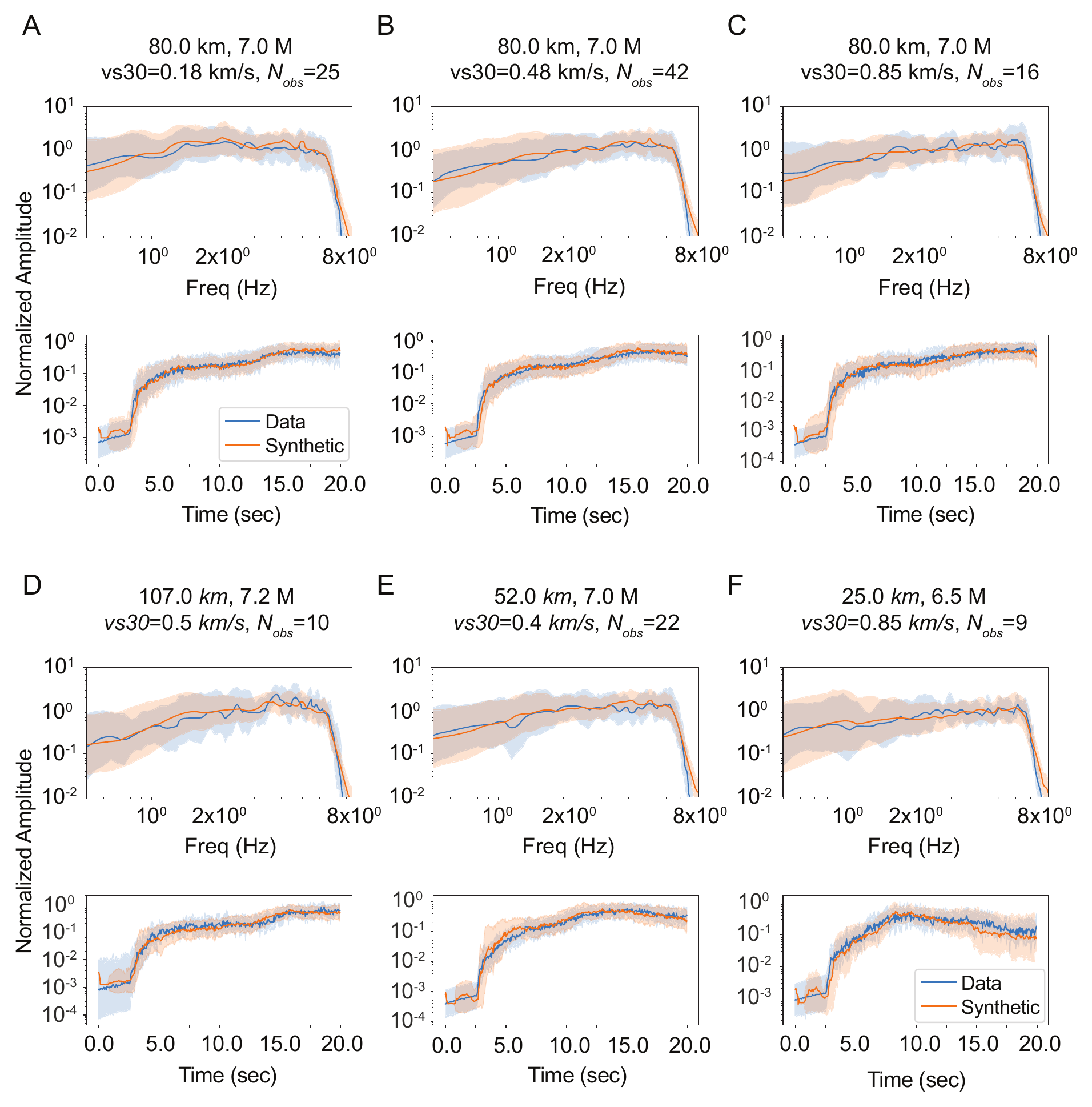}
    \caption{
    Assessment of the quality of our best conditional GAN model trained using three conditional variables: Event-Station distance, magnitude, and $Vs30$. In each panel, average Normalized amplitude spectra (upper pane) and average normalized acceleration envelopes (lower pane) are used to contrast real (blue) and synthetic (orange) acceleration time-histories. The title of each panel contains the bin-midpoints used to synthesize accelerograms. $N_{obs}$ is the number of real observations used to compute statistical averages. }
    \label{fig:vs30_mag_dist}
\end{figure}

To compare our model predictions with statistical quantities computed on real accelerograms, we divide our dataset into discrete magnitude-distance-vs30 bins. We synthesize $N=256$ waveforms using the bin mid-points as inputs and, whenever possible, randomly select the same number of real accelerograms in the given bin for comparison. The generator was trained using continuous conditional variables as inputs; thus, an ideal testing design would require extremely narrow bins. (Figure \ref{fig:vs30_mag_dist}). In practice, data availability limits the size of the bins. We used bin widths of $14 km$ in event-station distance and $0.2$ magnitude units. For $Vs30$,  we relied on available data to select bin widths. Meaningful statistical comparisons are possible when a sufficiently large number of observations are available in each bin, at least 20. However, given the nature of earthquakes statistics, and the current design of strong-motion sensor networks, it is challenging to gather observations of large earthquakes $(M > 6.0)$, especially at short distances $ D < 40 km $. Those cases are the most relevant for engineering applications; therefore, we were forced to compute TD and FD statistics using only a handful of real accelerograms in some situations. See Figure \ref{fig:vs30_mag_dist} for detailed comparisons in several representative frequency-magnitude-vs30 bins. 

\subsection{Peak Ground Acceleration}

As discussed in section \ref{section:data}, the generator produces normalized 3-Component waveforms and their corresponding normalization factors. We take the Peak Ground Acceleration (PGA) as the maximum of the three normalization constants.

\begin{figure}[htp]
    \centering
    \includegraphics[width=\textwidth]{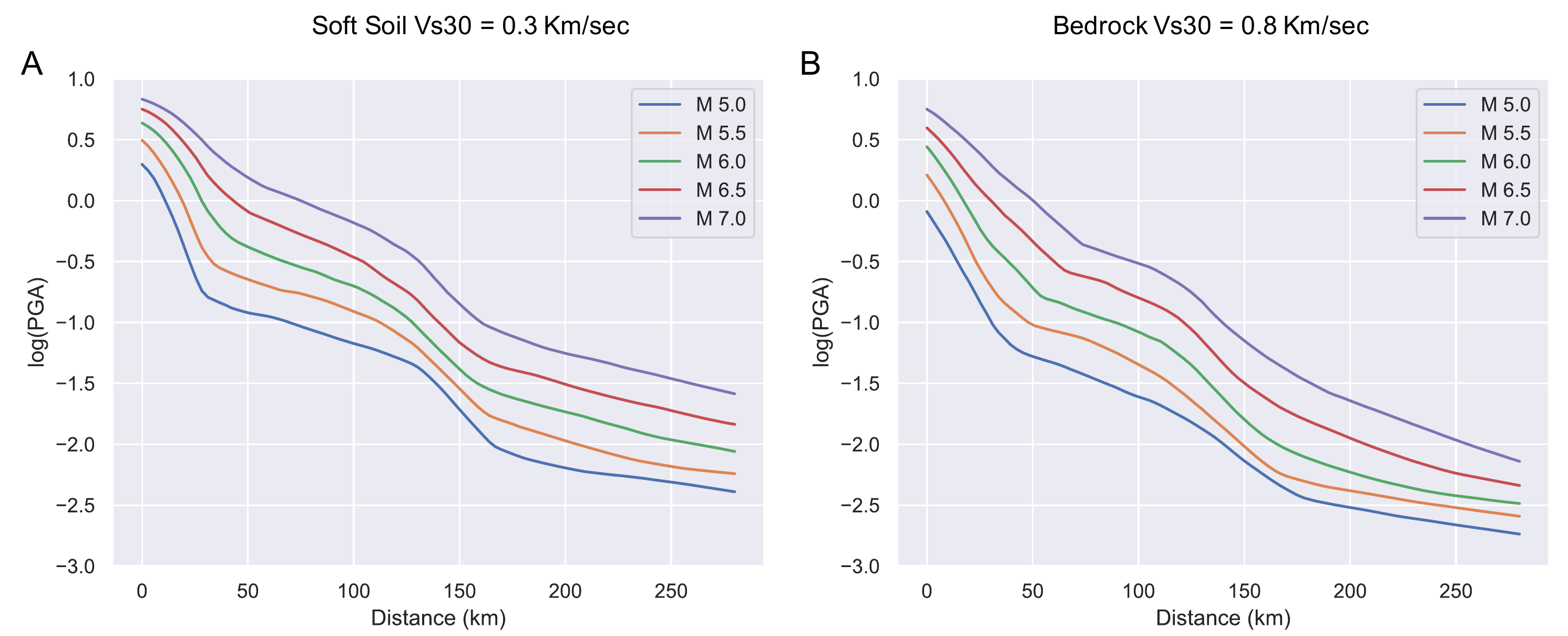}
    \caption{ Predicted Peak Ground Acceleration (PGA) as a function of distance for several representative magnitudes. (A) a soft soil site with $Vs30=0.3 km/sec$ and (B) a bedrock site with $Vs30=0.8 km/sec$.}
    \label{fig:pga_syn}
\end{figure}

\begin{figure}[htp]
    \centering
    \includegraphics[width=\textwidth]{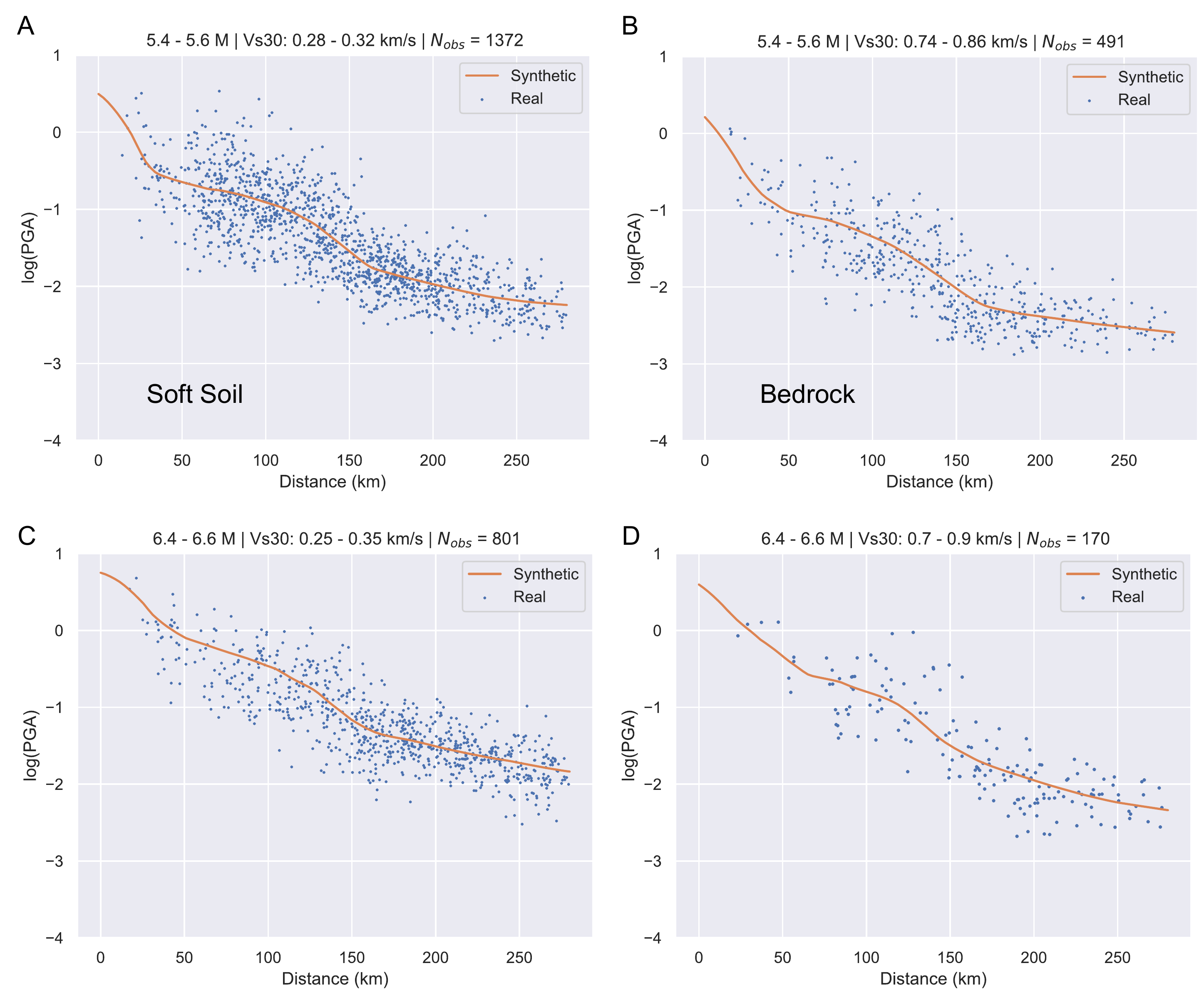}
    \caption{Synthetic and observed Peak Ground Acceleration (PGA) as a function of distance for different magnitude-vs30 bins. Each panel title contains bin edges and the number of observations, $N_{obs}$, in the bin. The continuous function (orange) was calculated using bin midpoints as inputs to our best generator model.  Blue dots correspond to observed PGAs.}
    \label{fig:pga_obs}
\end{figure}

In Figure \ref{fig:pga_syn}, we display $\log_{10}(PGA)$ as a function of distance for several representative magnitudes and fixed values of $Vs30$. We plot two panels, one for a soft site $(Vs30=0.3km_{sec})$  and other for hard bedrock $(Vs30=0.3km_{sec})$. As a byproduct of our seismogram generation strategy, we have built a model that has many of the essential ingredients of Ground Motion Prediction Equations (GMPEs).  Figure \ref{fig:pga_obs} displays physically reasonable, relatively smooth curves that capture the most relevant trends present in our dataset. Careful inspection of figure \ref{fig:pga_obs} reveals that our model could be used to calculate PGA in regions of the parameter space where no data is available. In the next section, we examine whether this is captured by the synthetic accelerograms, and the ability of the model to interpolate waveforms to conditions not previously observed.

\subsection{Interpolation Experiment}

\begin{figure}[htp]
    \centering
    \includegraphics[width=\textwidth]{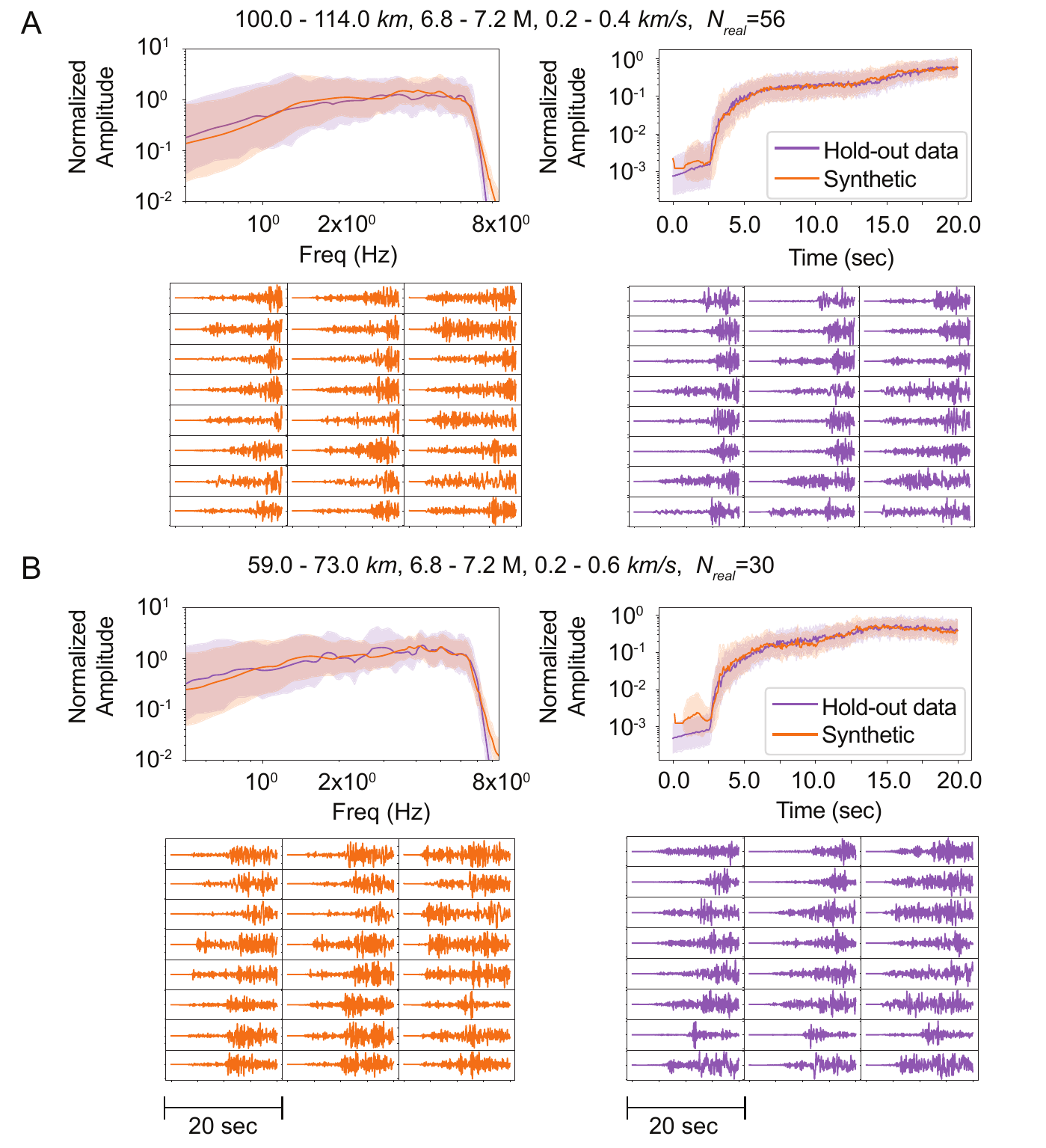}
    \caption{Results of two interpolation experiments. Each panel's title displays information about the range of values for which data was removed (Hold-out set). In orange, synthetic waveforms, in purple, real waveforms belonging to the Hold-out dataset. }
    \label{fig:interpolation}
\end{figure}

We design a series of experiments to determine whether the generator model can be used to interpolate acceleration time histories to conditions where no earthquake ground motion recordings exist. We select a distance-magnitude-vs30 bin from the original dataset for each experiment, containing at least 30 observations and having a distance width of no more than $30 km$. Figure's \ref{fig:interpolation} experiments focus on bins with large magnitude earthquakes, Supp. Figure \ref{fig:supp_3vc_val} concentrates on bins with event-station distances less than $60 km$. We remove all accelerograms belonging to the selected bin and retrain our model. The removed data is held out for subsequent testing. Once the generator model is trained, we use the bin mid-points to generate $N=256$ synthetic seismograms. We then compute average Fourier amplitude spectra, and average acceleration envelops for the synthetic and holdout sets. Finally, we calculate standard deviations in log-space and overlay the corresponding results in the same plot (Figure \ref{fig:interpolation}), which allows us to visually assess the generator's model quality.

We perform an initial series of interpolation experiments using only two conditional variables:  distance and magnitude (Supp. Figure \ref{fig:supp_2vc_val}). These two variables control most of the first-order features observed in accelerograms, e.g., the time difference between the first onset (P-wave) and the second one (S-wave) directly correlates with distance. Also, larger magnitude events have amplitude spectra with relatively more energy concentrated at lower frequencies, making visual validation of the results straight forward. 

Distance and Magnitude are the essential ingredients of any ground motion prediction model. Hence, as a methodological strategy, we started by building a robust model using only these two variables and extended it by introducing an additional variable: Vs30, a proxy for site response. In Figure \ref{fig:interpolation} and Supp. Figure \ref{fig:supp_3vc_val}, we show the results of our interpolation experiments run using our full model with its three conditional variables: distance, magnitude, and Vs30.

\subsection{Training Stability and Hyperparameter tuning}
\label{section:model_sel}
A critical advantage of WGANs is that they provide a useful metric that correlates well with the quality of the generator output, the Wasserstein distance $W(\Pg,\Preal)$ as approximated by the discriminator loss \citep{arjovsky_towards_2017}:

\begin{equation} \label{eq:w_loss}
W_{D} =  \E_{x \sim \Preal}[D(x)] - \E_{z \sim p}[ D(G(z)) ]
\end{equation}

Note that these are simply the first two terms in equation \ref{eq:loss_disc}, but with the sign flipped, because $W(\Pg,\Preal)$ is defined as a maximum, and during the optimization process we search for a minimum. 

We perform a series of experiments to assess the sensitivity of the training process to different hyperparameter choices. In each experiment, we split our initial data into two randomly selected non-overlapping sets: 80\% for training and 20 \% for validation. We adversarially train our models about 160 epochs and compute  $W_{D}$ on the training set after each generator iteration. The validation loss is computed after a full epoch. The training loss is a continuous snapshot of the optimization process, so it looks noisier; the validation loss, on the other hand, is a smooth average that allows us to assess the stability and the generalizability of our model.

\begin{figure}[htp]
    \centering
    \includegraphics[width=\textwidth]{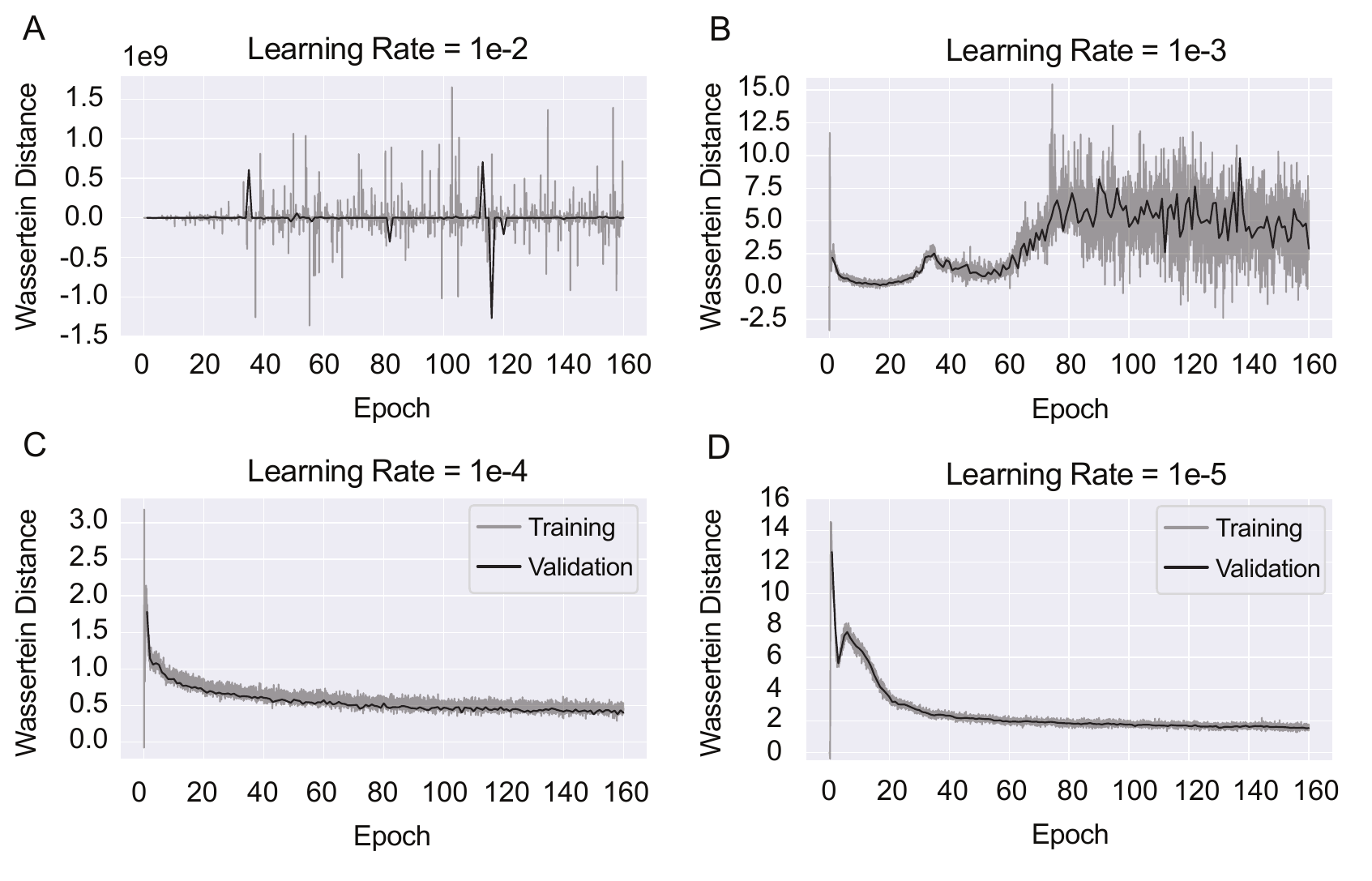}
    \caption{ Wasserstein distance for different learning rates. }
    \label{fig:lr_exp}
\end{figure}

In Figure \ref{fig:lr_exp}, we show the results of four experiments where we vary the learning rate.   Following \citep{gulrajani_improved_2017} we used an Adam optimizer \citep{kingma_adam_2014} with $\beta_1=0.0$ and $\beta_2=0.9$ . We choose a learning rate of  $1e-4$, much lower than the default because it gives stable results, and, among all experiments performed, it results in the smallest  $W_{D}$ value. Equation \ref{eq:w_loss} provides a useful metric to diagnose potential issues.

\begin{figure}[htp]
    \centering
    \includegraphics{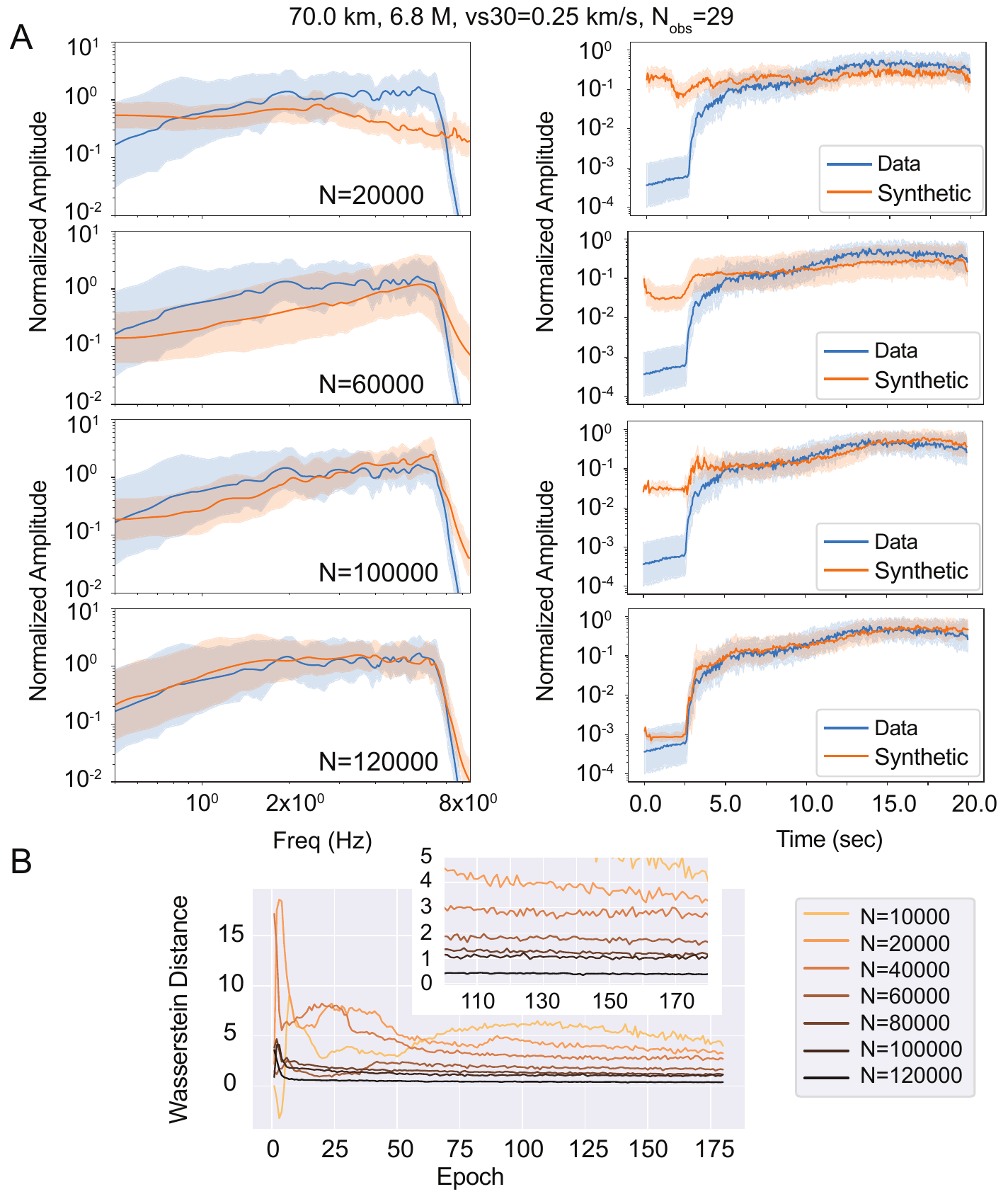}
    \caption{ (A) Average normalized amplitude spectra (left column), average normalized acceleration envelopes (right column); each row displays the results of an experiment using a dataset constructed by selecting N observations at random from the original data.  (B) Curves of Wasserstein Distance convergence for different experiments. Figure title contains the bin-midpoints used to synthesize accelerograms and $N_{obs}$, the number of real accelerograms in the bin, used for comparison. }
    \label{fig:samp_size}
\end{figure}

We also explore the effect of sample size (Figure \ref{fig:samp_size}), by randomly selecting subsets of $N$ observations from our original dataset. For $N<=40000$, we observed no signs of convergence; instead, the validation and training losses tend to plateau at relatively high values of $W_{D}$; this behavior correlates with poor quality of the generated accelerograms, as shown in the first row of Figure \ref{fig:samp_size}. At $N=60,000$ output quality starts to improve, but even when using $N=100,000$ samples the trained model is not able to capture important details of the accelerogram's envelop distribution. We conclude that at least $N=120,000$ samples are needed for our model to converge to realistically looking waveforms. These sample size experiments are specific to our dataset, but can serve as a guideline when applying our method to other regions.

\section{Discussion and Conclusion}
Generating realistic ground motion time histories for hypothetical earthquakes remains challenging. Deterministic simulation methods are promising but not yet practical. We have taken advantage of the increasing availability of strong-motion sensor data and recent advances in machine learning to propose a completely alternative approach. We developed an adversarial training scheme that allowed a generator deep neural network $G$ to learn the conditional probability distribution of a massive ground motion data set from Japan. Our results suggest that $G$ approximates well the many complex processes that give rise to the observed earthquake ground motions, the acceleration time series it synthesizes have all the essential ingredients of real accelerograms in both time and frequency domains (Figure \ref{fig:vs30_mag_dist} and Figure \ref{fig:cwaves}). 

An advantage of our technique is that it directly captures site effects. Simulation-based methods typically model ground motions for reference bedrock sites; the resulting acceleration spectrum has to be modified in an ad-hoc fashion by multiplying it with empirically derived site-specific amplification factors. Our model synthesizes acceleration times series with Fourier amplitude spectral shapes that closely follow the ones of real accelerograms for different $Vs30$ values (Figure \ref{fig:vs30_mag_dist} ).  Interestingly, there is consistently more energy at lower frequencies (0.1-2 Hz) for smaller values of $vs30$ ($0.1-0.3 km/s$). Our results confirm that vs30 is an excellent proxy for site response. It exerts an important control on Fourier amplitude spectral shapes; this is apparent from the data and the synthesized accelerograms (Figure \ref{fig:vs30_mag_dist} ).

A second distinguishing feature of our methodology is that it is stochastic by design. It does not produce a single waveform consistent with a set of input parameters; it provides a full range of possible ground motions that accurately reflect the data's variability.  There are two critical reasons for this: (1) $G$
 learns to approximate a conditional probability distribution, and (2) our framework never attempts to fit the data in the time or the frequency domains. The adversarial training scheme allows the discriminator model to provide a dynamic measure of similarity that continuously improves. Thus, the synthesized acceleration time histories reflect the variability of potential ground motions in the region where the data was collected. Such a synthetic set of ground motions may allow engineering to perform dynamical analysis of structures in a Bayesian sense and to determine robust confidence intervals for their estimates.

As with any machine learning approach, the computational cost is paid upfront during training.  Our model takes about 8 hours to train on an Nvidia Tesla V100 GPU, but once the process is complete, it can synthesize hundreds of accelerograms in seconds, even when running on modest laptop hardware. We provide a pre-trained generator model and open-source code so that any user can quickly synthesize ground motions of interest.  Deterministic simulation of a single scenario earthquake has a computational cost orders of magnitude larger. It requires specialized High-Performance Computing (HPC) hardware not available to most users, and it is limited by the lack of detailed knowledge of the earth's structure and rupture characteristics.

Our approach lays a foundation for data-driven synthesis of accelerograms, and we envision additional work in the future expanding on this framework. We have chosen a simple, functional model architecture for this study: 400 output units per channel, limiting us to 20-second long waveforms sampled at $20 Hz$ and three conditional input variables. We have proven that it works well for a wide range of scenario earthquakes in Japan. We have also provided all the necessary ingredients and guidance to adapt it to any other region's data set.  The conditional variables were chosen to reflect the key controlling factors of earthquake ground motions: Distance from the hypocenter, event size, and site response. Since our framework is stochastic, it may not be necessary or even desirable to include additional variables. The generator model already provides a full set of accelerograms representing many faulting styles, source characteristics, and relevant hypocentral depths. Furthermore, if needed, our framework can be combined with a suitable set of Ground Motion Prediction Equations (GMPEs); the model does provide reliable PGA estimates (Figure \ref{fig:pga_syn}  and Figure \ref{fig:pga_obs}), but it would be straight forward to use the normalized output waveforms along with PGA or PGV estimates obtained using regional GMPEs.

While our model has been shown to successfully interpolate waveforms given conditioning variables not seen before, its main limitation is that it cannot extrapolate outside of this range. Users of the trained model should be aware of this and plan subsequent usage of the model accordingly. For our dataset, the main concerns here are the high and low ends of the magnitude spectrum. As additional data become available in the future, it may be possible to expand this range. An additional minor challenge of our method currently is that in a handful of cases, our model struggles to approximate the first 2.5 seconds of noise before the p-wave onset. Inspection of envelope statistics of real accelerograms (Figure \ref{fig:vs30_mag_dist}) reveals that pre-signal noise variability is often large, spanning many orders of magnitude. Therefore, the noise may come from different distributions, varying across stations, with a complex signature that is challenging to capture. 

Longer accelerograms, with a potentially more accurate noise signature, can be synthesized by increasing the number of layers and the number of output units. Larger models tend to perform better but would take significantly longer to train. Since we had to iterate over a large number of network architectures and hyperparameter choices in a reasonable time, we kept our model as small as possible; nevertheless, given sufficient computing power, the same architecture can be readily scaled up.

The waveforms the generator model can synthesize are already appropriate for many engineering applications. We are confident that the framework we have developed is robust and well-suited for many ground-motion prediction tasks. Further work is still needed, but the goal of synthesizing on-demand, accurate acceleration time-histories for any scenario earthquake might be within reach.

\renewcommand{\figurename}{Supp. Figure}
\renewcommand{\thefigure}{1}
\begin{figure}[htp]
    \centering
    \includegraphics[width=\textwidth]{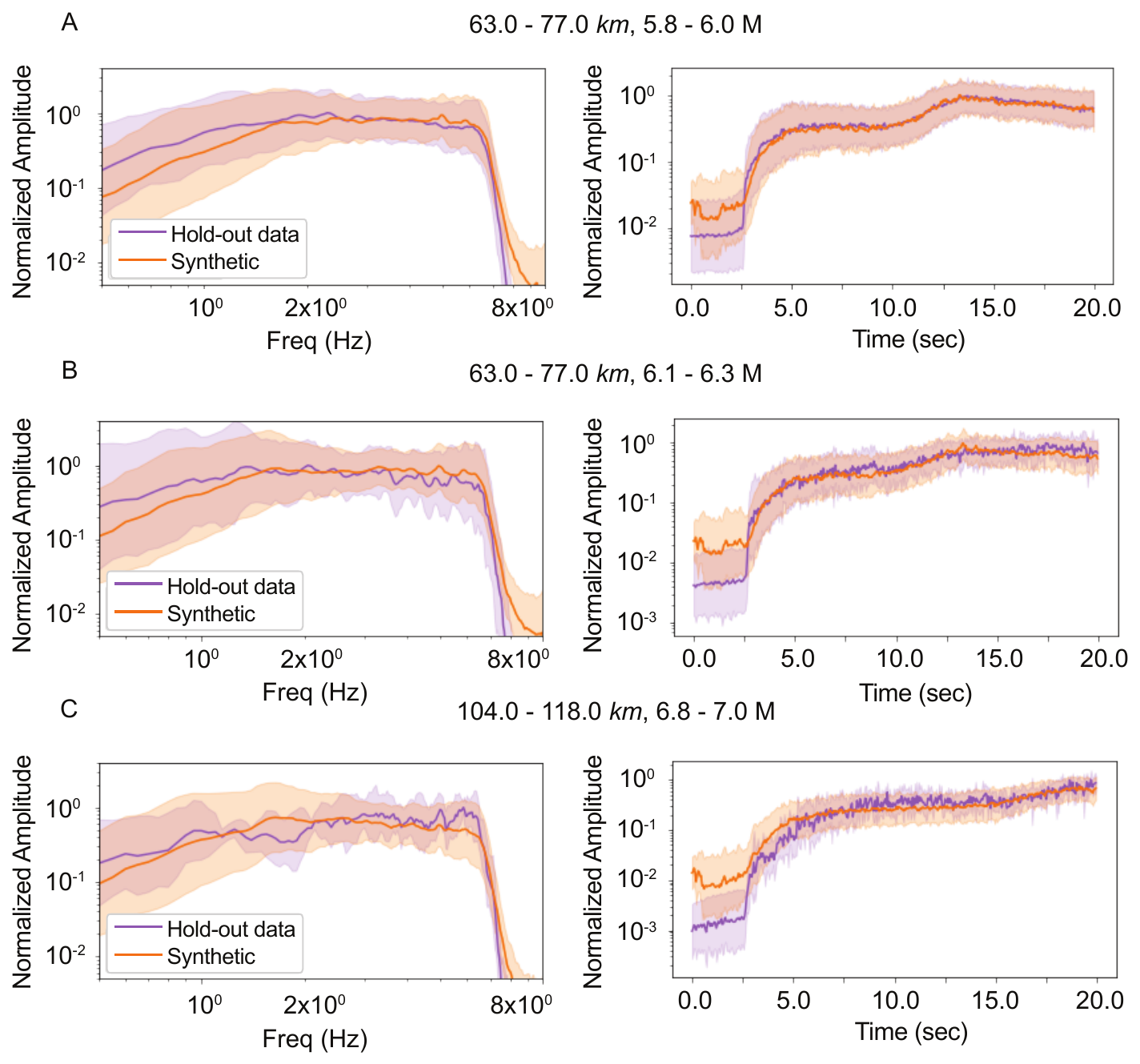}
    \caption{Results of interpolation experiments using only two conditional variables: Event-Station distance and Magnitude.}
    \label{fig:supp_2vc_val}
\end{figure}

\renewcommand{\thefigure}{2}
\begin{figure}[htp]
    \centering
    \includegraphics[width=\textwidth]{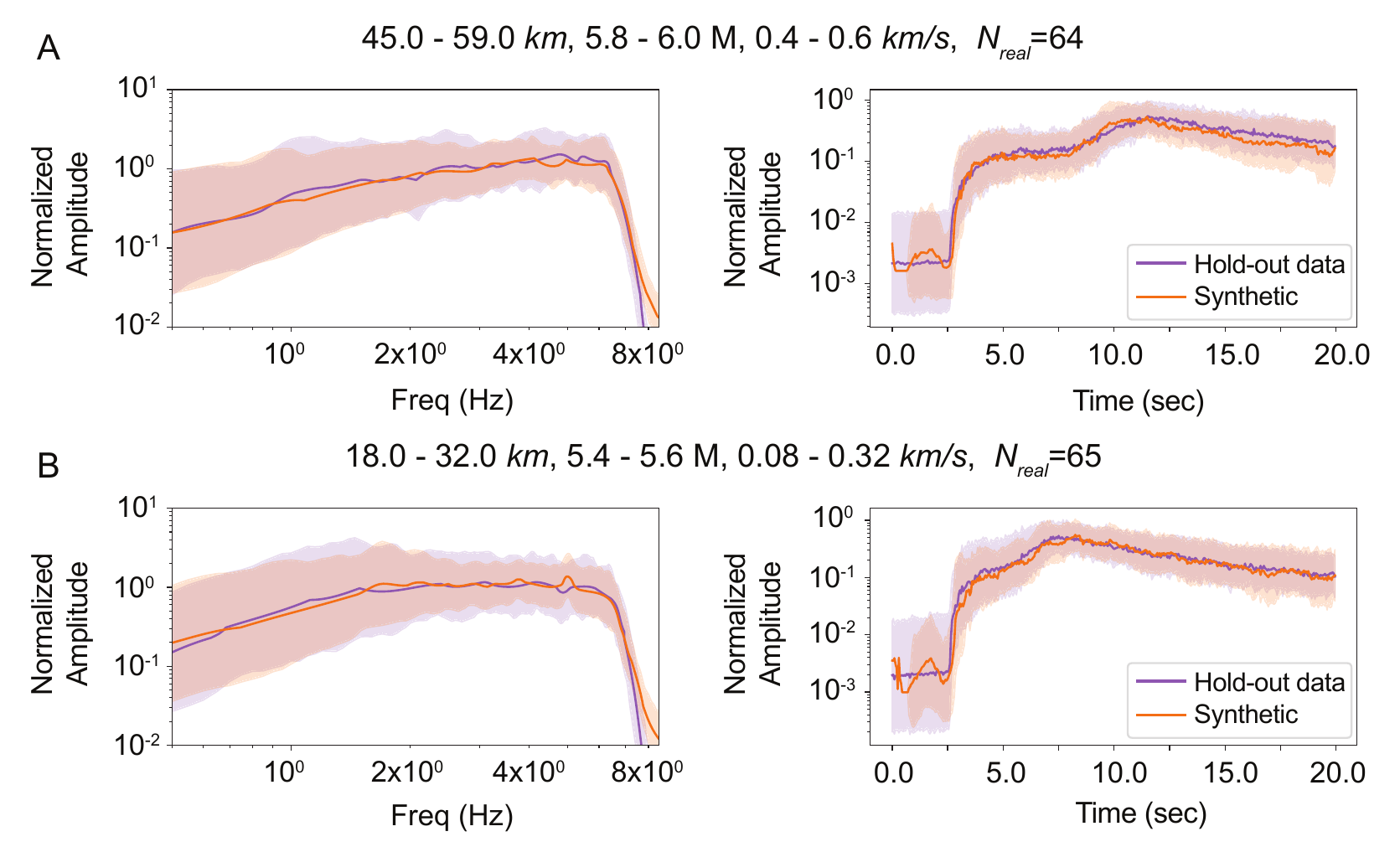}
    \caption{Additional results for interpolation experiments using three conditional variables: Event-Station distance, Magnitude, Vs30.}
    \label{fig:supp_3vc_val}
\end{figure}

\printbibliography  

@article{hancock_improved_2006,
	title = {An Improved Method of Matching Response Spectra of Recorded Earthquake Ground Motion Using Wavelets},
	volume = {10},
	issn = {1363-2469},
	url = {https://doi.org/10.1080/13632460609350629},
	doi = {10.1080/13632460609350629},
	pages = {67--89},
	issue = {sup001},
	journaltitle = {Journal of Earthquake Engineering},
	author = {Hancock, Jonathan and Watson-Lamprey, jennie and Abrahamson, Norman and Bommer, Julian J. and Markatis, Alexandros and {McCoyh}, Emma and Mendis, Rishmila},
	urldate = {2020-08-12},
	date = {2006-01-01},
}

@article{hancock_numbers_2008,
	title = {Numbers of scaled and matched accelerograms required for inelastic dynamic analyses},
	volume = {37},
	rights = {Copyright © 2008 John Wiley \& Sons, Ltd.},
	issn = {1096-9845},
	doi = {10.1002/eqe.827},
	pages = {1585--1607},
	number = {14},
	journaltitle = {Earthquake Engineering \& Structural Dynamics},
	author = {Hancock, Jonathan and Bommer, Julian J. and Stafford, Peter J.},
	urldate = {2020-08-12},
	date = {2008},
	langid = {english},
}

@article{mena_pseudodynamic_2012,
	title = {Pseudodynamic Source Characterization for Strike‐Slip Faulting Including Stress Heterogeneity and Super‐Shear {RupturesPseudodynamic} Source Characterization for Strike‐Slip Faulting Including Stress Heterogeneity},
	volume = {102},
	issn = {0037-1106},
	url = {https://pubs.geoscienceworld.org/ssa/bssa/article/102/4/1654/325384/Pseudodynamic-Source-Characterization-for-Strike},
	doi = {10.1785/0120110111},
	pages = {1654--1680},
	number = {4},
	journaltitle = {Bulletin of the Seismological Society of America},
	shortjournal = {Bulletin of the Seismological Society of America},
	author = {Mena, B. and Dalguer, L. A. and Mai, P. M.},
	urldate = {2020-08-15},
	date = {2012-08-01},
	langid = {english},
	note = {Publisher: {GeoScienceWorld}},
}

@article{boore_simulation_2003-1,
	title = {Simulation of Ground Motion Using the Stochastic Method},
	volume = {160},
	issn = {1420-9136},
	url = {https://doi.org/10.1007/PL00012553},
	doi = {10.1007/PL00012553},
	pages = {635--676},
	number = {3},
	journaltitle = {pure and applied geophysics},
	shortjournal = {Pure appl. geophys.},
	author = {Boore, D. M.},
	urldate = {2020-08-13},
	date = {2003-03-01},
	langid = {english},
}

@article{kaul_spectrum-consistent_1978,
	title = {Spectrum-Consistent Time-History Generation},
	volume = {104},
	url = {https://cedb.asce.org/CEDBsearch/record.jsp?dockey=0008198},
	pages = {781--788},
	number = {4},
	journaltitle = {Journal of the Engineering Mechanics Division},
	author = {Kaul, Maharaj K.},
	urldate = {2020-08-13},
	date = {1978},
	note = {Publisher: {ASCE}},
}

@article{naeim_use_1995,
	title = {On the Use of Design Spectrum Compatible Time Histories},
	volume = {11},
	issn = {8755-2930},
	url = {https://doi.org/10.1193/1.1585805},
	doi = {10.1193/1.1585805},
	pages = {111--127},
	number = {1},
	journaltitle = {Earthquake Spectra},
	shortjournal = {Earthquake Spectra},
	author = {Naeim, Farzad and Lew, Marshall},
	urldate = {2020-08-13},
	date = {1995-02-01},
	langid = {english},
	note = {Publisher: {SAGE} Publications Ltd {STM}},
}

@book{gasparini_simulated_1976,
	location = {Cambridge, {MA}},
	title = {Simulated earthquake motions compatible with prescribed response spectra},
	publisher = {Massachusetts Institute of Technology, Dept. of Civil Engineering, Constructed Facilities Division},
	author = {Gasparini, Dario A and Vanmarcke, Erik},
	date = {1976},
	note = {{OCLC}: 1176024695},
}

@article{wang_seismogen_2019,
	title = {{SeismoGen}: Seismic Waveform Synthesis Using Generative Adversarial Networks},
	url = {http://arxiv.org/abs/1911.03966},
	shorttitle = {{SeismoGen}},
	journaltitle = {{arXiv}:1911.03966 [physics, stat]},
	author = {Wang, Tiantong and Trugman, Daniel and Lin, Youzuo},
	urldate = {2020-08-13},
	date = {2019-11-10},
	eprinttype = {arxiv},
	eprint = {1911.03966},
	note = {version: 1},
}

@article{komatitsch_spectral_1998,
	title = {The spectral element method: An efficient tool to simulate the seismic response of 2D and 3D geological structures},
	volume = {88},
	issn = {0037-1106},
	url = {https://pubs.geoscienceworld.org/ssa/bssa/article/88/2/368/120304/The-spectral-element-method-An-efficient-tool-to},
	shorttitle = {The spectral element method},
	pages = {368--392},
	number = {2},
	journaltitle = {Bulletin of the Seismological Society of America},
	shortjournal = {Bulletin of the Seismological Society of America},
	author = {Komatitsch, Dimitri and Vilotte, Jean-Pierre},
	urldate = {2020-08-12},
	date = {1998-04-01},
	langid = {english},
	note = {Publisher: {GeoScienceWorld}},
}

@article{herrero_kinematic_1994,
	title = {A kinematic self-similar rupture process for earthquakes},
	volume = {84},
	issn = {0037-1106},
	url = {https://pubs.geoscienceworld.org/ssa/bssa/article/84/4/1216/119869/A-kinematic-self-similar-rupture-process-for},
	pages = {1216--1228},
	number = {4},
	journaltitle = {Bulletin of the Seismological Society of America},
	shortjournal = {Bulletin of the Seismological Society of America},
	author = {Herrero, A. and Bernard, P.},
	urldate = {2020-08-12},
	date = {1994-08-01},
	langid = {english},
	note = {Publisher: {GeoScienceWorld}},
}

@article{graves_simulating_1996,
	title = {Simulating seismic wave propagation in 3D elastic media using staggered-grid finite differences},
	volume = {86},
	issn = {0037-1106},
	url = {https://pubs.geoscienceworld.org/ssa/bssa/article/86/4/1091/120141/Simulating-seismic-wave-propagation-in-3D-elastic},
	pages = {1091--1106},
	number = {4},
	journaltitle = {Bulletin of the Seismological Society of America},
	shortjournal = {Bulletin of the Seismological Society of America},
	author = {Graves, Robert W.},
	urldate = {2020-08-12},
	date = {1996-08-01},
	langid = {english},
	note = {Publisher: {GeoScienceWorld}},
}

@article{ma_effects_2007,
	title = {Effects of Large-Scale Surface Topography on Ground Motions, as Demonstrated by a Study of the San Gabriel Mountains, Los Angeles, {CaliforniaEffects} of Large-Scale Surface Topography on Ground Motions},
	volume = {97},
	issn = {0037-1106},
	url = {https://pubs.geoscienceworld.org/ssa/bssa/article/97/6/2066/341892/Effects-of-Large-Scale-Surface-Topography-on},
	doi = {10.1785/0120070040},
	pages = {2066--2079},
	number = {6},
	journaltitle = {Bulletin of the Seismological Society of America},
	shortjournal = {Bulletin of the Seismological Society of America},
	author = {Ma, Shuo and Archuleta, Ralph J. and Page, Morgan T.},
	urldate = {2020-08-12},
	date = {2007-12-01},
	langid = {english},
	note = {Publisher: {GeoScienceWorld}},
}

@article{hanks_character_1981,
	title = {The character of high-frequency strong ground motion},
	volume = {71},
	issn = {0037-1106},
	url = {https://pubs.geoscienceworld.org/ssa/bssa/article/71/6/2071/102118/The-character-of-high-frequency-strong-ground},
	pages = {2071--2095},
	number = {6},
	journaltitle = {Bulletin of the Seismological Society of America},
	shortjournal = {Bulletin of the Seismological Society of America},
	author = {Hanks, Thomas C. and {McGuire}, Robin K.},
	urldate = {2020-08-12},
	date = {1981-12-01},
	langid = {english},
	note = {Publisher: {GeoScienceWorld}},
}

@article{heaton_estimating_1986,
	title = {Estimating ground motions using recorded accelerograms},
	volume = {8},
	issn = {1573-0956},
	url = {https://doi.org/10.1007/BF01904051},
	doi = {10.1007/BF01904051},
	pages = {25--83},
	number = {1},
	journaltitle = {Surveys in Geophysics},
	shortjournal = {Surv Geophys},
	author = {Heaton, Thomas H. and Tajima, Fumiko and Mori, Ann Wildenstein},
	urldate = {2020-08-12},
	date = {1986-03-01},
	langid = {english},
}

@article{bommer_use_2004,
	title = {The use of real earthquake accelerograms as input to dynamic analysis},
	volume = {08},
	issn = {1363-2469},
	url = {https://www.worldscientific.com/doi/10.1142/S1363246904001596},
	doi = {10.1142/S1363246904001596},
	pages = {43--91},
	issue = {spec01},
	journaltitle = {Journal of Earthquake Engineering},
	shortjournal = {J. Earth. Eng.},
	author = {Bommer, Julian J. and Acevedo, Ana Beatriz},
	urldate = {2020-08-12},
	date = {2004-01-01},
	note = {Publisher: Imperial College Press},
}

@article{brune_tectonic_1970,
	title = {Tectonic stress and the spectra of seismic shear waves from earthquakes},
	volume = {75},
	rights = {Copyright 1970 by the American Geophysical Union.},
	issn = {2156-2202},
	url = {https://agupubs.onlinelibrary.wiley.com/doi/abs/10.1029/JB075i026p04997},
	doi = {10.1029/JB075i026p04997},
	pages = {4997--5009},
	number = {26},
	journaltitle = {Journal of Geophysical Research (1896-1977)},
	author = {Brune, James N.},
	urldate = {2020-08-12},
	date = {1970},
	langid = {english},
	note = {\_eprint: https://agupubs.onlinelibrary.wiley.com/doi/pdf/10.1029/{JB}075i026p04997},
}

@article{boore_stochastic_1983,
	title = {Stochastic simulation of high-frequency ground motions based on seismological models of the radiated spectra},
	volume = {73},
	issn = {0037-1106},
	url = {https://pubs.geoscienceworld.org/ssa/bssa/article/73/6A/1865/118579/Stochastic-simulation-of-high-frequency-ground},
	pages = {1865--1894},
	number = {6},
	journaltitle = {Bulletin of the Seismological Society of America},
	shortjournal = {Bulletin of the Seismological Society of America},
	author = {Boore, David M.},
	urldate = {2020-08-12},
	date = {1983-12-01},
	langid = {english},
	note = {Publisher: {GeoScienceWorld}},
}

@article{graves_broadband_2010,
	title = {Broadband Ground-Motion Simulation Using a Hybrid {ApproachBroadband} Ground-Motion Simulation Using a Hybrid Approach},
	volume = {100},
	issn = {0037-1106},
	url = {https://pubs.geoscienceworld.org/ssa/bssa/article/100/5A/2095/325180/Broadband-Ground-Motion-Simulation-Using-a-Hybrid},
	doi = {10.1785/0120100057},
	pages = {2095--2123},
	number = {5},
	journaltitle = {Bulletin of the Seismological Society of America},
	shortjournal = {Bulletin of the Seismological Society of America},
	author = {Graves, Robert W. and Pitarka, Arben},
	urldate = {2020-08-12},
	date = {2010-10-01},
	langid = {english},
	note = {Publisher: {GeoScienceWorld}},
}

@article{prieto_fortran_2009,
	title = {A Fortran 90 library for multitaper spectrum analysis},
	volume = {35},
	issn = {0098-3004},
	url = {http://www.sciencedirect.com/science/article/pii/S0098300409000077},
	doi = {10.1016/j.cageo.2008.06.007},
	pages = {1701--1710},
	number = {8},
	journaltitle = {Computers \& Geosciences},
	shortjournal = {Computers \& Geosciences},
	author = {Prieto, G. A. and Parker, R. L. and Vernon {III}, F. L.},
	urldate = {2020-07-01},
	date = {2009-08-01},
	langid = {english},
}

@inproceedings{borcherdt_vs30_2012,
	title = {{VS}30 – A site-characterization parameter for use in building Codes, simplified earthquake resistant design, {GMPEs}, and {ShakeMaps}},
	url = {https://pubs.er.usgs.gov/publication/70041709},
	eventtitle = {The 15th World Conference on Earthquake Engineering},
	author = {Borcherdt, Roger D.},
	urldate = {2020-06-04},
	date = {2012},
}

@article{power_overview_2008,
	title = {An Overview of the {NGA} Project},
	volume = {24},
	issn = {8755-2930},
	url = {https://doi.org/10.1193/1.2894833},
	doi = {10.1193/1.2894833},
	pages = {3--21},
	number = {1},
	journaltitle = {Earthquake Spectra},
	shortjournal = {Earthquake Spectra},
	author = {Power, Maurice and Chiou, Brian and Abrahamson, Norman and Bozorgnia, Yousef and Shantz, Thomas and Roblee, Clifford},
	urldate = {2020-06-04},
	date = {2008-02-01},
	langid = {english},
	note = {Publisher: {SAGE} Publications Ltd {STM}},
}

@article{douglas_survey_2008,
	title = {A Survey of Techniques for Predicting Earthquake Ground Motions for Engineering Purposes},
	volume = {29},
	issn = {1573-0956},
	url = {https://doi.org/10.1007/s10712-008-9046-y},
	doi = {10.1007/s10712-008-9046-y},
	pages = {187},
	number = {3},
	journaltitle = {Surveys in Geophysics},
	shortjournal = {Surv Geophys},
	author = {Douglas, John and Aochi, Hideo},
	urldate = {2020-06-04},
	date = {2008-10-10},
	langid = {english},
}

@inproceedings{ioffe_batch_2015,
	location = {Lille, France},
	title = {Batch normalization: accelerating deep network training by reducing internal covariate shift},
	series = {{ICML}'15},
	shorttitle = {Batch normalization},
	pages = {448--456},
	booktitle = {Proceedings of the 32nd International Conference on International Conference on Machine Learning - Volume 37},
	publisher = {{JMLR}.org},
	author = {Ioffe, Sergey and Szegedy, Christian},
	urldate = {2020-06-03},
	date = {2015-07-06},
}

@article{odena_deconvolution_2016,
	title = {Deconvolution and Checkerboard Artifacts},
	volume = {1},
	issn = {2476-0757},
	url = {http://distill.pub/2016/deconv-checkerboard},
	doi = {10.23915/distill.00003},
	pages = {e3},
	number = {10},
	journaltitle = {Distill},
	shortjournal = {Distill},
	author = {Odena, Augustus and Dumoulin, Vincent and Olah, Chris},
	urldate = {2020-06-03},
	date = {2016-10-17},
	langid = {english},
}

@article{mirza_conditional_2014,
	title = {Conditional Generative Adversarial Nets},
	url = {http://arxiv.org/abs/1411.1784},
	journaltitle = {{arXiv}:1411.1784 [cs, stat]},
	author = {Mirza, Mehdi and Osindero, Simon},
	urldate = {2020-06-03},
	date = {2014-11-06},
	eprinttype = {arxiv},
	eprint = {1411.1784},
}

@inproceedings{reed_generative_2016,
	title = {Generative Adversarial Text to Image Synthesis},
	url = {http://proceedings.mlr.press/v48/reed16.html},
	eventtitle = {International Conference on Machine Learning},
	pages = {1060--1069},
	booktitle = {International Conference on Machine Learning},
	author = {Reed, Scott and Akata, Zeynep and Yan, Xinchen and Logeswaran, Lajanugen and Schiele, Bernt and Lee, Honglak},
	urldate = {2020-06-03},
	date = {2016-06-11},
	langid = {english},
	note = {{ISSN}: 1938-7228
Section: Machine Learning},
}

@article{li_machine_2018,
	title = {Machine Learning Seismic Wave Discrimination: Application to Earthquake Early Warning},
	volume = {45},
	rights = {©2018. American Geophysical Union. All Rights Reserved.},
	issn = {1944-8007},
	url = {https://agupubs.onlinelibrary.wiley.com/doi/abs/10.1029/2018GL077870},
	doi = {10.1029/2018GL077870},
	shorttitle = {Machine Learning Seismic Wave Discrimination},
	pages = {4773--4779},
	number = {10},
	journaltitle = {Geophysical Research Letters},
	author = {Li, Zefeng and Meier, Men-Andrin and Hauksson, Egill and Zhan, Zhongwen and Andrews, Jennifer},
	urldate = {2020-06-02},
	date = {2018},
	langid = {english},
}

@book{villani_optimal_2009,
	location = {Berlin Heidelberg},
	title = {Optimal Transport: Old and New},
	isbn = {978-3-540-71049-3},
	url = {https://www.springer.com/gp/book/9783540710493},
	series = {Grundlehren der mathematischen Wissenschaften},
	shorttitle = {Optimal Transport},
	publisher = {Springer-Verlag},
	author = {Villani, Cédric},
	urldate = {2020-06-02},
	date = {2009},
	langid = {english},
	doi = {10.1007/978-3-540-71050-9},
}

@article{saito_tganv2_2018,
	title = {{TGANv}2: Efficient Training of Large Models for Video Generation with Multiple Subsampling Layers},
	url = {http://arxiv.org/abs/1811.09245},
	shorttitle = {{TGANv}2},
	journaltitle = {{arXiv}:1811.09245 [cs]},
	author = {Saito, Masaki and Saito, Shunta},
	urldate = {2020-06-01},
	date = {2018-11-22},
	eprinttype = {arxiv},
	eprint = {1811.09245},
}

@inproceedings{donahue_adversarial_2018,
	title = {Adversarial Audio Synthesis},
	url = {https://openreview.net/forum?id=ByMVTsR5KQ},
	eventtitle = {International Conference on Learning Representations},
	author = {Donahue, Chris and {McAuley}, Julian and Puckette, Miller},
	urldate = {2020-06-01},
	date = {2018-09-27},
}

@article{karras_progressive_2018,
	title = {Progressive Growing of {GANs} for Improved Quality, Stability, and Variation},
	url = {http://arxiv.org/abs/1710.10196},
	journaltitle = {{arXiv}:1710.10196 [cs, stat]},
	author = {Karras, Tero and Aila, Timo and Laine, Samuli and Lehtinen, Jaakko},
	urldate = {2020-06-01},
	date = {2018-02-26},
	eprinttype = {arxiv},
	eprint = {1710.10196},
}

@article{alain_what_2014,
	title = {What regularized auto-encoders learn from the data-generating distribution},
	volume = {15},
	issn = {1532-4435},
	pages = {3563--3593},
	number = {1},
	journaltitle = {The Journal of Machine Learning Research},
	shortjournal = {J. Mach. Learn. Res.},
	author = {Alain, Guillaume and Bengio, Yoshua},
	date = {2014-01-01},
}

@inproceedings{bengio_better_2013,
	title = {Better Mixing via Deep Representations},
	url = {http://proceedings.mlr.press/v28/bengio13.html},
	eventtitle = {International Conference on Machine Learning},
	pages = {552--560},
	booktitle = {International Conference on Machine Learning},
	author = {Bengio, Yoshua and Mesnil, Gregoire and Dauphin, Yann and Rifai, Salah},
	urldate = {2020-06-01},
	date = {2013-02-13},
	langid = {english},
	note = {{ISSN}: 1938-7228
Section: Machine Learning},
}

@article{bengio_representation_2013,
	title = {Representation Learning: A Review and New Perspectives},
	volume = {35},
	issn = {1939-3539},
	doi = {10.1109/TPAMI.2013.50},
	shorttitle = {Representation Learning},
	pages = {1798--1828},
	number = {8},
	journaltitle = {{IEEE} Transactions on Pattern Analysis and Machine Intelligence},
	author = {Bengio, Yoshua and Courville, Aaron and Vincent, Pascal},
	date = {2013-08},
	note = {Conference Name: {IEEE} Transactions on Pattern Analysis and Machine Intelligence},
}

@inproceedings{ledig_photo-realistic_2017,
	title = {Photo-Realistic Single Image Super-Resolution Using a Generative Adversarial Network},
	doi = {10.1109/CVPR.2017.19},
	eventtitle = {2017 {IEEE} Conference on Computer Vision and Pattern Recognition ({CVPR})},
	pages = {105--114},
	booktitle = {2017 {IEEE} Conference on Computer Vision and Pattern Recognition ({CVPR})},
	author = {Ledig, Christian and Theis, Lucas and Huszár, Ferenc and Caballero, Jose and Cunningham, Andrew and Acosta, Alejandro and Aitken, Andrew and Tejani, Alykhan and Totz, Johannes and Wang, Zehan and Shi, Wenzhe},
	date = {2017-07},
	note = {{ISSN}: 1063-6919},
}

@article{radford_unsupervised_2015,
	title = {Unsupervised representation learning with deep convolutional generative adversarial networks},
	journaltitle = {{arXiv} preprint {arXiv}:1511.06434},
	author = {Radford, Alec and Metz, Luke and Chintala, Soumith},
	date = {2015},
}

@inproceedings{salimans_improved_2016,
	title = {Improved techniques for training gans},
	pages = {2234--2242},
	booktitle = {Advances in neural information processing systems},
	author = {Salimans, Tim and Goodfellow, Ian and Zaremba, Wojciech and Cheung, Vicki and Radford, Alec and Chen, Xi},
	date = {2016},
}

@inproceedings{isola_image--image_2017,
	title = {Image-to-image translation with conditional adversarial networks},
	pages = {1125--1134},
	booktitle = {Proceedings of the {IEEE} conference on computer vision and pattern recognition},
	author = {Isola, Phillip and Zhu, Jun-Yan and Zhou, Tinghui and Efros, Alexei A.},
	date = {2017},
}

@article{kingma_adam_2014,
	title = {Adam: A method for stochastic optimization},
	shorttitle = {Adam},
	journaltitle = {{arXiv} preprint {arXiv}:1412.6980},
	author = {Kingma, Diederik P. and Ba, Jimmy},
	date = {2014},
}

@article{arjovsky_towards_2017,
	title = {Towards Principled Methods for Training Generative Adversarial Networks},
	url = {http://arxiv.org/abs/1701.04862},
	journaltitle = {{arXiv}:1701.04862 [cs, stat]},
	author = {Arjovsky, Martin and Bottou, Léon},
	urldate = {2020-05-30},
	date = {2017-01-17},
	eprinttype = {arxiv},
	eprint = {1701.04862},
}

@inproceedings{arjovsky_wasserstein_2017,
	title = {Wasserstein Generative Adversarial Networks},
	url = {http://proceedings.mlr.press/v70/arjovsky17a.html},
	eventtitle = {International Conference on Machine Learning},
	pages = {214--223},
	booktitle = {International Conference on Machine Learning},
	author = {Arjovsky, Martin and Chintala, Soumith and Bottou, Léon},
	urldate = {2020-05-30},
	date = {2017-07-17},
	langid = {english},
	note = {{ISSN}: 1938-7228
Section: Machine Learning},
}

@article{gulrajani_improved_2017,
	title = {Improved Training of Wasserstein {GANs}},
	url = {http://arxiv.org/abs/1704.00028},
	journaltitle = {{arXiv}:1704.00028 [cs, stat]},
	author = {Gulrajani, Ishaan and Ahmed, Faruk and Arjovsky, Martin and Dumoulin, Vincent and Courville, Aaron},
	urldate = {2020-05-29},
	date = {2017-12-25},
	eprinttype = {arxiv},
	eprint = {1704.00028},
}

@incollection{goodfellow_generative_2014,
	title = {Generative Adversarial Nets},
	url = {http://papers.nips.cc/paper/5423-generative-adversarial-nets.pdf},
	pages = {2672--2680},
	booktitle = {Advances in Neural Information Processing Systems 27},
	publisher = {Curran Associates, Inc.},
	author = {Goodfellow, Ian and Pouget-Abadie, Jean and Mirza, Mehdi and Xu, Bing and Warde-Farley, David and Ozair, Sherjil and Courville, Aaron and Bengio, Yoshua},
	editor = {Ghahramani, Z. and Welling, M. and Cortes, C. and Lawrence, N. D. and Weinberger, K. Q.},
	urldate = {2020-05-29},
	date = {2014},
}
\end{document}